# Investigation of microinverter based on the two-switch DC-DC flyback converter topology

El Iysaouy Lahcen[1, 4], Edvardas Bielskis[2], Lahbabi Mhammed[1], Algirdas Baskys[3], Oumnad Abdelmajid[4]

[1]The Signals, Systems and Components Laboratory, Sidi Mohamed Ben Abdellah University FST Fez, Morocco

[2]Center for Physical Sciences and Technology,
Sauletekio al.3, LT-10257 Vilnius, Lithuania

[3]Department of Computer Engineering, Vilnius Gediminas Technical University,
Naugarduko St. 41, LT-03227 Vilnius, Lithuania

[4]Department of Electrical Engineering, Mohamed V University EMI Rabat, Morocco

[1] lahcen.eliysaouy@usmba.ac.ma

*Abstract*—**The investigation results of the grid connected photovoltaic single stage microinverter with a new topology are presented. The microinverter is based on the couple of two-switch DC-DC flyback converters and has simple design. The investigation results show that the analyzed microinverter is characterized by the high efficiency and low total harmonic distortion of output voltage.**

## I. INTRODUCTION

Due to their non-sustainability and harmful effects on the environment, fossil fuels no longer stand as attractive sources of energy. Solar radiation is a significant source of renewable energy and is likely to constitute a major source of future energy. Moreover, solar cells are unique in that they directly convert the incident solar irradiation into electricity. Thus, photovoltaic (PV) power management concepts are essential to extract as much power as possible from the solar energy. PV energy systems are being extensively studied because of its benefits of environmental friendly and renewable characteristics.

Several PV systems are used different configurations to achieve high DC voltage output[1], [2]. These configurations include following connection methods: parallel series (SP), total-cross-tied (TCT), Honey Comb (HC), bridge linked (BL) and Sudo-ku[1], [3]–[5]. However, the PV panels often work in mismatching conditions due to different panel orientations and shadowing effects. This mismatching problem can cause serious problems for the power-conditioning system. To overpowered this drawback, several publications has proposed a module-integrated converter concept[6], [7]. Such a PV systems composed of a PV panels with an individual Photovoltaic DC-AC inverters are called PV flyback microinverters[7]–[11]. For the module-integrated flyback microinverter on a PV panel, a long operation lifespan and low power loss must be ensured. The output filter is a critical component of the microinverter to improve the output power quality and reduce total harmonic distortion (THD). This is because the output current of the system contains high-frequency harmonic caused by the pulse width modulation (PWM) switching. For the voltage source inverter (VSI), the simplest output line filter only has one inductor. For the better performance of a filter, the CL and LCL filter can be configured with combinations of capacitors and inductors [12]–[15].

## II. THE MICROINVERTER WITH THE TWO-SWICH DC-DC FLYBACK CONVERTER TOPOLOGY

*Operation Principle*

Figure 1 shows the topology of investigated single stage microinverter based on the couple of two-switch DC-DC flyback converters [16]. The primary side circuit consists of power switches (M1, M2, M3, M4) and a transformer T.
The secondary-side circuit consists of switches (M5, M6) and a filter CL. U is the input voltage. $U_{ac}$ is the output voltage grid connected. The positive half period of output voltage is formed using two-switch DC-DC flyback converter based on transistors M1 and M4 (transistors M2 and M3 are in state OFF) and the output voltage is forwarded to the grid thought the transistor M5 which is in state ON. The negative half period of output voltage is formed using two-switch DC-DC flyback converter based on transistors M2 and M3 (transistors M1 andM4 are in state OFF) and the output voltage is delivered to the grid through the transistor M6. Then the high frequency harmonics of the formed AC voltage caused by the switching of converter switches are filtered by the low pass filter. The converter output voltage waveforms and spectrum and converter efficiency was investigated. The investigation was performed for two low pass filter types: LCL and CL. However, in this paper we will focus just on CL filter.

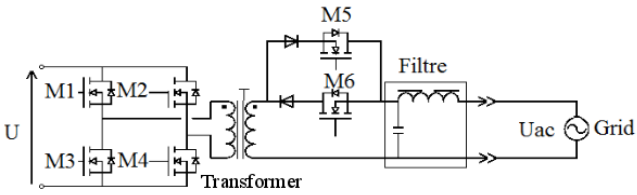

Fig. 1: Topology of microinverter based on the couple of two-switch DC-DC flyback converters

### III. OUTPUT FILTER

The high switching frequency provides high-order harmonics to output current[15]. Therefore, an output filter is required for the rejection of switching frequency components. CL filter is widely used for current source inverter (CSI) and LCL filter which has an additional inductor component to the conventional CL filter, is proposed in [16]. The benefits of LCL filter are appeared on the system size and the components capacity. Although, we will focus on the characteristics of CL filter. The investigation was performed for 50 Hz microinverter output voltage frequency 25 kHz switching frequency. Generally, a CL filter is used to couple the current source inverter (CSI) to the voltage source grid. For the attenuation of switching frequency components, the CL filter is designed as an output filter. Figure 1 shows the circuit of the microinverter topology with the CL filter[17]. As the grid impedance remains unknown, to set up the transfer function of the circuit and design parameters, the additional inductive reactance (Lg) term is defined and used with the CL filter parameters (L and Cf) as shown in Figure 2. The performance of the CL filter depends on the position of the resonant frequency of the filter. Therefore, the key of designing parameters is that the frequency of harmonics that have to be filtered would be behind the resonant frequency of the filter[17]. The transfer function for the CL filter is expressed a

$$A_{LC} = \frac{U_g}{U_c} = \frac{1}{S^2 CL + 1}.$$

where Uc is the voltage side of flyback converter and Ug is the output voltage side of grid connected of the filter CL. According to the control theory of the second-order system, the damped frequency is equal to $fc = \frac{1}{2\pi\sqrt{LC}}$

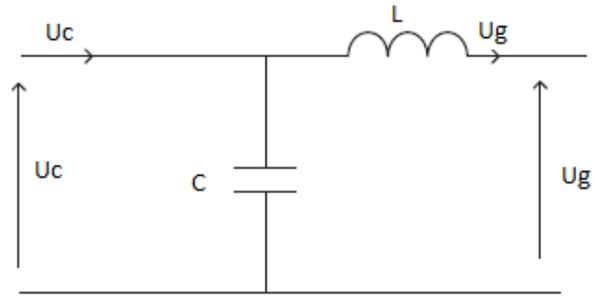

Fig. 2: Circuit diagram of CL Filter

### IV. SIMULATION RESULTS

The investigation was performed using simulation based on the software Matlab/Simulink. The microconverter output voltage waveforms and spectrum and converter efficiency were investigated. The obtained output voltage waveforms and spectrum are illustrated in Figure 3(a, b, c). The results show that the Total Harmonic Distortion (THD) of the output voltage for different voltage amplitude (amp) at 25000 Hz switching frequency (Sf) is not higher than 5% and respect the requirements of IEEE standard 1574. The results show that the low pass (CL) filter plays an important role in grid-connected converter when trying to reduce switching-frequency voltage ripples injected into the grid.

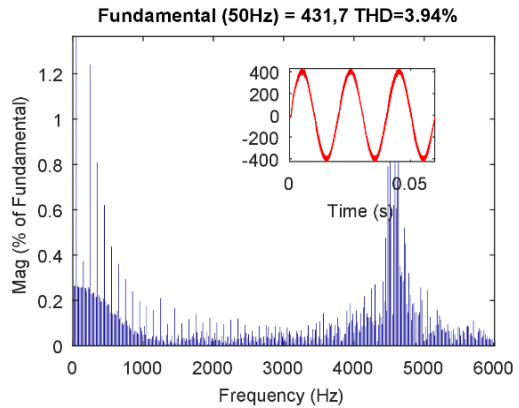

(a) THD of the output voltage for amp (400 RMS).

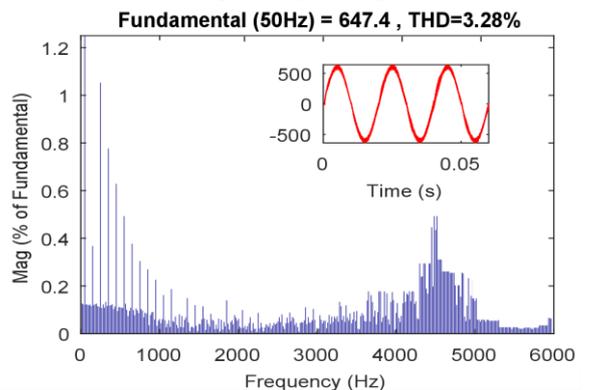

(b) THD of the output voltage for amp (600 RMS).

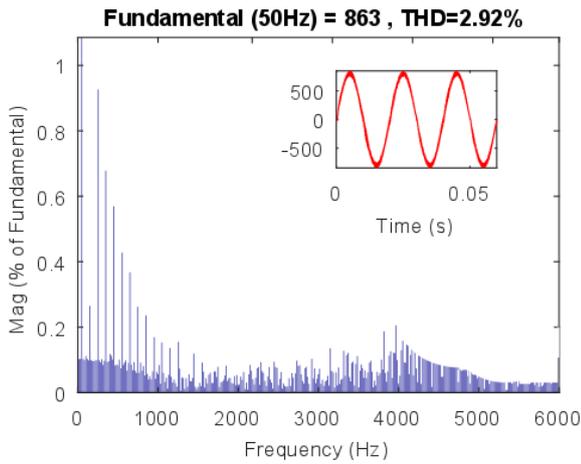

(c) THD of the output voltage for amp (800 RMS).
Fig. 3: The spectrum and Total Harmonic distortion THD of the output voltage at switching frequency 25000 Hz

The obtained output voltage waveforms and spectrum are illustrated in Figure 3. The results showed that the THD of the output voltage at 25000 Hz switching frequency is not higher than 4% and respect the requirements of IEEE.

The dependences of investigated single stage microinverter efficiency on output power at various internal resistances of flyback converter switches M1-M4 in state ON are presented on Figure 4. It is seen that the efficiency of converter decreases when internal resistance of switch transistors in state ON increases

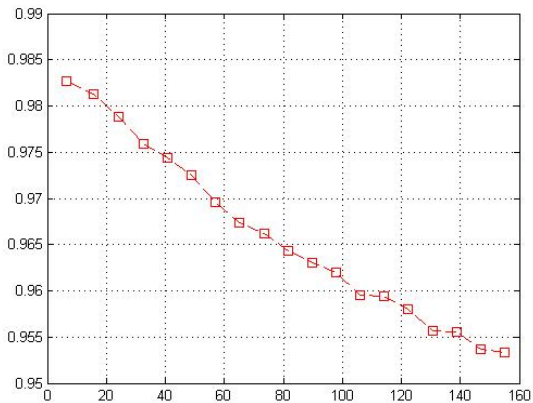

(a)  Ron = 0.008 $\Omega$

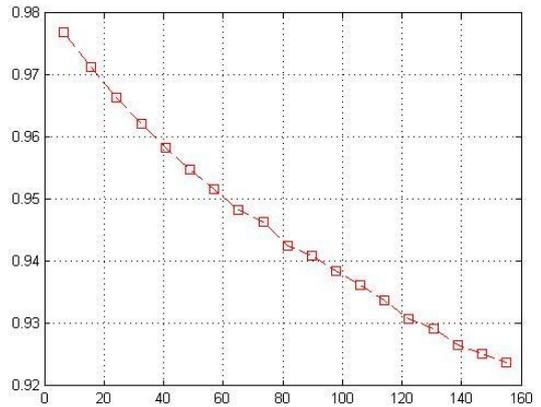

(b)  Ron = 0.04 $\Omega$

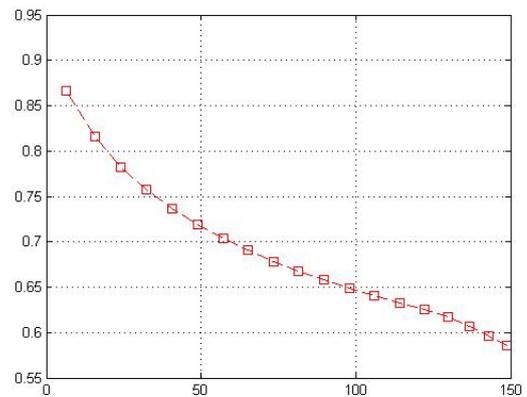

(c)  Ron = 0.5 $\Omega$
Fig. 4: Dependences of microconverter efficiency on output power at various internal resistances of switch transistors M1-M4 in state ON

## V. CONCLUSION

In this The results of investigation of microinverter based on the couple of two-switch DC-DC flyback converters show that the THD of the output voltage at 25000 Hz is not higher than 4% and respect the requirements of IEEE. Furthermore, the efficiency of microconverter reaches 0.96 at 100W output power at 25 kHz switching frequency when the internal resistance of MOSFET switches M1-M4 are around 0.008 $\Omega$.

Finally, it can be conclude that the investigated grid connected microinverter Is characterized by the high efficiency and provides low THD of output voltage.

ACKNOWLEDGMENT

This work was supported by Vilnius Gediminas Technical University Lithuania.
.